\documentclass{article}

\usepackage{arxiv}

\usepackage[utf8]{inputenc} 
\usepackage[T1]{fontenc}    
\usepackage{hyperref}       
\usepackage{url}            
\usepackage{booktabs}       
\usepackage{amsfonts}       
\usepackage{nicefrac}       
\usepackage{microtype}      
\usepackage{lipsum}		
\usepackage{amssymb,amsmath}
\usepackage{listings}
\usepackage{graphicx}
\usepackage{subfig}

\title{Data assimilation in Agent-based models using creation and annihilation operators}

\author{
  Daniel Tang\\
  Leeds Institute for Data Analytics\\
  University of Leeds\\
  Leeds, UK\\
  \texttt{D.Tang@leeds.ac.uk} \\
}

\begin{document}
\maketitle

\begin{abstract}
Agent-based models are a powerful tool for studying the behaviour of complex systems that can be described in terms of multiple, interacting ``agents''. However, because of their inherently discrete and often highly non-linear nature, it is very difficult to reason about the relationship between the state of the model, on the one hand, and our observations of the real world on the other. In this paper we consider agents that have a discrete set of states that, at any instant, act with a probability that may depend on the environment or the state of other agents. Given this, we show how the mathematical apparatus of quantum field theory can be used to reason probabilistically about the state and dynamics the model, and describe an algorithm to update our belief in the state of the model in the light of new, real-world observations. Using a simple predator-prey model on a 2-dimensional spatial grid as an example, we demonstrate the assimilation of incomplete, noisy observations and show that this leads to an increase in the mutual information between the actual state of the observed system and the posterior distribution given the observations, when compared to a null model.
\end{abstract}

\keywords{Data assimilation, Agent based model, Quantum field theory, Probabilistic programming, Fock space}

\section{Introduction}

When modelling real-world phenomena we often use the model to make forecasts, only to be later faced with new observations that weren't available at the time of the forecast. This leads to the problem of how to ``assimilate'' the new observations into an updated forecast and, perhaps, another forecast for a later time. There exists an extensive literature on this problem as applied to weather forecasting models\cite{kalnay2003atmospheric}\cite{carrassi2018data} which describes how the problem can be reduced using Bayes' rule:
\begin{equation}
P(s|\Omega) \propto P(\Omega|s)P(s)
\label{Bayes}
\end{equation}
Here, $P(s)$ is our original forecast (this is a probability distribution since there is uncertainty in the forecast) $P(\Omega|s)$ is the likelihood of the observation, $\Omega$, given a forecast, $s$, and $P(s|\Omega)$ is our updated forecast. However, the algorithms developed to do this in weather forecasting models depend on the model's governing equations having some degree of smoothness, a condition not generally satisfied by agent-based models.

If there are only a very small number of observations, particle filtering or Markov Chain Monte Carlo methods may have some success\cite{russell2009artificial}, but as the number of observations increases, these methods tend to quickly fail. This is because as the number of observations increases, the proportion of the model's phase-space that has non-zero likelihood often shrinks exponentially, so if the dimensionality of the phase-space is not trivially small, finding a non-zero start point for a Markov chain becomes exponentially harder. Similarly for particle filters; the number of particles needed to prevent all particles ending up with zero probability, increases exponentially with the number of observations.

For these reasons, the problem of data assimilation in agent-based models remains an unsolved problem, and this imposes a severe restriction on the utility of these models when applied to real world problems.

\section{Quantum field theory applied to agent-based models}

In order to apply Bayes' rule to agent-based models (ABMs) we must first devise a way to represent a probability distribution over the possible states of an agent-based model. The task is complicated somewhat by the fact that, in general, the exact number of agents in the model is unknown and variable, so the set of model states must span models with any number of agents. Luckily, we can use the mathematical formalism of quantum field theory to do this. In a quantum field there are an unknown number of interacting particles and the quantum field is represented as a quantum amplitude over the possible configurations of particles; to apply this to ABMs, we just need to interpret the \textit{particles} as agents and replace the \textit{quantum amplitudes} with probabilities.

\subsection{The creation operator}
Let $\emptyset$ represent the empty model state (i.e. the state with no agents present), and let $a^\dag_i$ be an operator on model states that has the effect of adding an agent in state $i$ to the model state, this is called the \textit{creation operator}. So, for example, $a_i^\dag\emptyset$ is the state with a singe agent in state $i$. Since the result of a creation operator is itself a state, we can apply another creation operator, for example, $a_j^\dag a_i^\dag\emptyset$ to represent a state with two agents, one in state $i$ and one in state $j$, or even $a_i^{\dag 2}\emptyset$ to represent a state with two agents in state $i$. We can go on applying operators in this way to construct any desired state.

If we now allow states to be multiplied by real numbers and added together, a vector space is induced which can represent probability distributions over model states. For example $0.5a_i^\dag\emptyset + 0.5a_j^\dag\emptyset$ would represent a probability distribution with $0.5$ probability that there's a single agent in state $i$ and $0.5$ probability that there's a single agent in state $j$ (with no probability that both agents are present at the same time).

Since each term in a probability distribution always ends in the empty state, $\emptyset$, we can just as easily consider the additions and multiplications to be part of the operator, so the example distribution above could be equivalently written $0.5(a_i^\dag + a_j^\dag)\emptyset$. Given this, composure of operators works in the expected way, for example, $a_z^\dag\left( 0.5(a_i^\dag + a_j^\dag)\right)\emptyset = 0.5(a_z^\dag a_i^\dag + a_z^\dag a_j^\dag)\emptyset$.

With this representation, we need a way to express the fact that it doesn't matter in what order the creation operators are applied, so $a_j^\dag a_i^\dag \emptyset = a_i^\dag a_j^\dag \emptyset$. Since this is true for all states, not just $\emptyset$, we can drop the $\emptyset$ and say\footnote{to be correct, the 0 is interpreted as the ``multiply by zero'' operator or alternatively $0I$ where $I$ is the identity operator}
\[
a_i^\dag a_j^\dag - a_j^\dag a_i^\dag = 0
\]
This form of equation is known as a \textit{commutation relation} or \textit{commutator} and has a shorthand notation:
\begin{equation}
[a_i^\dag, a_j^\dag] = 0
\end{equation}

\subsection{The annihilation operator}

Now let $a_i$ be the \textit{annihilation operator} which has the following properties:
\begin{equation}
a_i\emptyset = 0
\label{annihilation0}
\end{equation}
\begin{equation}
[a_i,a_j] = 0
\end{equation}
and
\begin{equation}
[a_i,a_j^\dag] = \delta_{ij}
\label{annihilationcommute}
\end{equation}
where $\delta_{ij}$ is the Kronecker delta function.

Given just these properties, we can show that
\[
a_ia_i^{\dag n}\emptyset = a_ia_i^\dag a_i^{\dag n-1}\emptyset = (a_i^\dag a_i + [a_i,a_i^\dag])a_i^{\dag n-1}\emptyset
\]
This is a recurrence relationship which, using equations \ref{annihilation0} and \ref{annihilationcommute}, can be solved to give
\[
a_ia_i^{\dag n}\emptyset = na_i^{\dag n-1}\emptyset
\]
So, the annihilation operator has the effect of multiplying by the number of agents in a state, then removing one of those agents.

Notice that, using the commutation relations in this way we can transform any sequence of operations on $\emptyset$ into an equivalent form that contains no annihilation operators. We'll call this the \textit{reduced form} of the operator.

\section{Equations of motion of a probabilistic ABM}

Armed with just the creation and annihilation operators we now show that if the behaviour of agents in an ABM can be described as a set of propensities to act and interact at any instant then we can transform the ABM into a ``Hamiltonian'' operator, containing only addition, multiplication, creation and annihilation operators. The Hamiltonian operator turns a probability distribution over ABM states into a rate of change of that distribution. This is significant in that it allows us to express the equations of motion of a probability distribution as an operator, and so describe probabilistically how the ABM evolves through time. That is, we've defined a ``probabilistic ABM'' that captures all the behaviour of the original ABM in a smooth, differentiable way which, as we'll show, makes it much easier to perform inference on the ABM.

To do this, start with the definition of an agent's behaviour: this must be representable as a set of actions and interactions. An action consists of a pair of functions, $(\rho(i), A(i))$ where $\rho$ is a rate per unit time and $A$ is a set of agents. An agent in state $i$ is said to have behaviour $(\rho(i), A(i))$ if in any infinitesimal time-slice, $\delta t$, it replaces itself by the agents in $A(i)$ with probability $\rho(i) \delta t$. An interaction consists of a pair of functions, $(\rho(i,j), A(i,j))$ such that an agent in state $i$, in the presence of another agent in state $j$ replaces both itself and the other agent with $A(i,j)$ with probability $\rho(i,j)\delta t$. Although this is not a common way to think about agent-based models, a very large class of models can easily be described in this way.

Without loss of generality, we assume all agents in a model have the same set of behaviours (different behaviours among agents can be achieved by putting an ``agent type'' marker in the agent's state and making the behaviour rates zero for certain agent types).

Consider now an action, $(\rho(i), A(i))$ and let $[k_{i1}\dots k_{in_i}] = A(i)$ be the results of the action. The Hamiltonian for this action would be
\[
H_{(\rho, A)} = \rho(i)\left(\prod_{k\in A(i)} a_k^\dag - a_i^\dag\right)a_i
\]
The product term expresses the rate of increase in probability in the state resulting from the action, while the $a_i^\dag$ term expresses the (equal in magnitude) rate of decrease in probability in the initial state due to the agent being removed when it acts. Because the annihilation operator, $a_i$, multiplies by the number of agents in a given state, the rate of change is also multiplied if there are multiple agents to which the behaviour applies (or multiplied by zero if there are no applicable agents).

Now consider an interaction $(\rho(i,j), I(i,j))$. The Hamiltonian for this interaction would be
\[
H_{(\rho, I)} = \rho(i,j)\left(\left(\prod_{k\in I(i,j)} a_k^\dag\right) - a_i^\dag a_j^\dag\right)a_ja_i
\]
The double annihilation operator, $a_ja_i$, has the effect of multiplying by the number of $i,j$ agent pairs so that each agent in state $i$ gets a chance to interact with each agent in state $j$. This also works when $j=i$, since $a_i^2a_i^{\dag n}\emptyset = n(n-1)a_i^{\dag n-2}\emptyset$, and the number of $i,i$ pairs is $n(n-1)$.

This pattern can easily be extended to interactions between three or more agents, but in this paper we'll consider only binary interactions.

If an agent has more than one behaviour, the Hamiltonian is simply the sum of the individual behaviours, so in this way we can build a Hamiltonian for a whole ABM. So, an ABM whose agents have a set of action behaviours $\left\{(\rho_1(i), A_1(i)) \dots (\rho_\alpha(i), A_\alpha(i))\right\}$ and a set of interaction behaviours $\left\{(\rho_1(i,j), I_1(i,j)) \dots (\rho_\beta(i,j), I_\beta(i,j))\right\}$ has the Hamiltonian
\[
H = \sum_{n=1}^\alpha\sum_i \rho_n(i)\left(\left(\prod_{k\in A_n(i)} a_k^\dag\right)  - a_i^\dag\right)a_i 
+ \sum_{m=1}^\beta \sum_i \sum_j \rho_m(i,j)\left(\left(\prod_{k\in I_m(i,j)} a_k^\dag\right) - a_i^\dag a_j^\dag\right)a_ja_i
\]
Although the number of terms in the Hamiltonian can become very large, we'll see later that we only ever need to deal computationally with the much smaller commutation relations $[X,H]$.

Given the Hamiltonian, the time evolution of a probability distribution over ABM states, $\psi$, is by definition
\[
\frac{d\psi}{dt} = H\psi
\]
This has the analytical solution
\begin{equation}
\psi_t = e^{tH}\psi_0
\label{timeoperator}
\end{equation}
where $\psi_0$ is the initial state, $\psi_t$ is the state at time $t$ and the exponential of an operator, $tH$, can be defined as
\[
e^{tH} = \sum_{n=0}^\infty \frac{t^nH^n}{n!}
\]

Equation \ref{timeoperator} gives us a prior forecast which can be inserted into Bayes' rule (equation \ref{Bayes}) in order to do data assimilation:
\begin{equation}
P(s_{t}|\Omega_{t}) = \frac{P(\Omega_{t}|s_{t})P(e^{tH}\psi_0 = s_{t})}{\sum_{s}P(\Omega_{t}|s)P(e^{tH}\psi_0 = s)}
\label{assimilation}
\end{equation}
where $s_{t}$ is a state at time $t$, $\Omega_{t}$ are the observations at time $t$ and $\psi_0$ is the probability distribution of the ABM at time $t=0$ given all observations up to that time.

So, we have a succinct, mathematical solution to the problem of data assimilation in an ABM. We now need to translate equation \ref{assimilation} into a numerical algorithm that will execute in a reasonable time.

\section{The Deselby distribution}

As a first step towards a practical numerical algorithm, we introduce the \textit{Deselby distribution} which we define as
\[
D_{i\lambda\Delta} = \sum_{k=0}^\infty \frac{(k)_\Delta \lambda^{k-\Delta} e^{-\lambda}}{k!} a_i^{\dag k}\emptyset
\]
where $\Delta$ is an integer, $\lambda$ is a real number and $(k)_\Delta$ denotes the $\Delta^{th}$-order falling factorial of $k$.

The Deselby distributions have the useful property that
\[
a_i^\dag D_{i\lambda\Delta} = D_{i\lambda(\Delta+1)}
\]
and so
\begin{equation}
a_i^{\dag \Delta}D_{i\lambda 0} = D_{i\lambda\Delta}
\label{deselbycreate}
\end{equation}

It can also be seen that
\begin{equation}
a_i D_{i\lambda0} = \lambda D_{i\lambda0}
\label{deselbyannihilate}
\end{equation}
Now imagine a product of Deselby distributions over all agent states
\[
D_\Lambda = \prod_i D_{i\lambda_i0}
\]
so that, for any $i$
\[
a_iD_\Lambda = \lambda_iD_\Lambda
\]
So, in the same way as with $\emptyset$, any sequence of operations on $D_\Lambda$ can be reduced to a sequence of creation operations on $D_\Lambda$, i.e. a reduced form, and we can consider $D_\Lambda$ as a ``ground'' state in just the way we have been using $\emptyset$ so far.

So, if we express a probability distribution in terms of operators on $D_\Lambda$ we can apply the Hamiltonian to this to get a rate of change in in terms of $D_\Lambda$. This will be much more computationally efficient than dealing directly with operators on $\emptyset$.

Note also that when $\Delta=0$, the Deselby distribution is just the Poisson distribution, so it is particularly convenient if we can express our prior beliefs about the number of agents in each state of an ABM in terms of Poisson distributions.

\subsection{Data assimilation with Deselby distributions}

As a concrete example, suppose there are a number of identical agents moving around a 2-dimensional grid, and we observe the number of agents in the $i^{th}$ grid-square. To make this slightly more general, suppose our method of observation may be imperfect and there is a probability, r, that an agent is detected given that it is there, such that the likelihood function is just the Binomial distribution
\[
P(m|k) = {k \choose m}r^m(1-r)^{k-m}
\]
where $m$ is the observed number of agents and $k$ is the actual number present. Appendix \ref{binomialderivation} shows that, if our prior is a Deselby distribution $D_{i\lambda\Delta}$, then the posterior is given by
\begin{equation}
P \propto  a_i^{\dag m}a_i^m Lg_{ir} D_{i\lambda\Delta}
\label{binomialposterior}
\end{equation}
Where we introduce the operator $Lg_{ir}$ which has the properties
\begin{equation}
[Lg_{ir}, a_i^\dag] = -ra_i^\dag Lg_{ir}
\end{equation}
\begin{equation}
[Lg_{ir}, a_i] = \frac{ra_i}{1-r} Lg_{ir}
\end{equation}
and
\begin{equation}
Lg_{ir}D_\Lambda = D_{\Lambda'}
\end{equation}
where $\Lambda'$ is $\Lambda$ with the $i^{th}$ component, $\lambda_i$, replaced by $(1-r)\lambda_i$.

Since euqation \ref{binomialposterior} is true for any Deselby distribution, it is also true for any state described as an operator on $D_\Lambda$. So, given $\psi_0$ (a distribution at time $t=0$ expressed as an operator on $D_\Lambda$) and an observation of $m$ agents in state $i$ at time $t$, the postrior distribution at time $t$ is given by
\begin{equation}
\psi_t \propto a_i^{\dag m}a_i^m Lg_{ir}e^{tH}\psi_0
\label{deselbyposterior}
\end{equation}

Multiple observations $\Omega = \left\{(i_1,m_1)...(i_n,m_n)\right\}$ can be treated by applying operators for each observation
\begin{equation}
\psi_t \propto \left(\prod_{(i,m)\in\Omega} a_i^{\dag m}a_i^m Lg_{ir}\right)\,e^{tH}\psi_0
\label{deselbyposterior}
\end{equation}

\subsection{Deselby state approximation}

Even when working with Deselby distributions, as time progresses the size of the reduced form of the distribution will become large and sooner or later. In order to maintain computational tractability, we'll need to somehow approximate the distribution. There are many ways of doing this, but a common way is to minimise the Kulback-Leibler divergence between the original distribution and the approximation\cite{murphy2012machine} over a family of possible approximations. This has the effect of minimising the loss of information due to the approximation.

If $P\emptyset$ is the original distribution and $A\emptyset$ the approximation, and we choose to minimise $D_{KL}(P\emptyset||A\emptyset)$ the resulting approximation has a particularly simple form.

By definition the KL-divergence is given by
\[
D_{KL}(P\emptyset||A\emptyset) = \emptyset\cdot P^* \left(\log(P\emptyset) - \log(A\emptyset)\right)
\]
where we define
\[
\log\left((c_1a^{\dag l_1} +  \dots + c_na^{\dag l_n})\emptyset\right) = (\log(c_1)a_1^{\dag l_1} +  \dots + \log(c_n)a_n^{\dag l_n})\emptyset
\]
and
\[
\emptyset \cdot a^{\dag n} \emptyset = \delta_{n0}
\]
where $\delta_{n0}$ is the Kronecker delta function.

$P^*$ is defined to be the \textit{conjugate} of $P$, where the conjugation operator is defined as having the properties
\[
(a_i^{\dag n})^* = \frac{a_i^n}{n!}
\]
\[
(A+B)^* = A^* + B^*
\]
and
\[
(AB)^* = A^*B^*
\]
and $c^* = c$ if $c$ is a constant.

If we choose $A$ to be a Deselby distribution, and minimise the KL divergence with respect to $\Lambda$ we have, for all $i$
\[
\forall_i: 0 = \emptyset\cdot P^*\frac{\partial \log(D_{i\lambda_i\Delta_i}\emptyset)}{\partial \lambda_i}
\]
Solving this for $\lambda_i$ gives

\begin{equation}
\forall_i:  \lambda_i = \emptyset \cdot e^{\sum_j a_j} a_iP\emptyset - \Delta_i
\label{mean}
\end{equation}

Where the operator
\[
\emptyset\cdot e^{\sum_j a_j} = \emptyset\cdot \prod_j\sum_{n=0}^\infty \frac{a_j^n}{n!}
\]
has the effect of summing over a distribution (i.e. summing the coefficients of an operator expressed in reduced form on $\emptyset$).

This can be interpreted as saying that the Deselby distribution, $D$, that minimises $D_{KL}(P\emptyset||D)$ is the one whose mean number of agents in state $i$ matches the mean of $P\emptyset$ for all $i$.

This gives us values for $\lambda_i$ given $\Delta_i$. To calculate $\Delta_i$, notice that for the KL-divergence to be finite, the support of $P\emptyset$ must be a subset of the support of $D$. So if $\Delta_i$ has value $n$, the probability that there are any less than $n$ agents in state $i$ in $P\emptyset$ must be zero. So, we can choose $\Delta_i$ to be the highest value for which there are definitely at least that many agents in state $i$ in $P\emptyset$.

\section{A practical algorithm for data assimilation}

Using equations \ref{deselbyposterior} and \ref{mean} we can perform a data assimilation cycle that consists of the following steps: take an initial state $\psi_0$, integrate forward in time, apply the set of observations $\Omega = \left\{(i_1, m_1)\dots(i_n,m_n)\right\}$, then finally approximate to the Deselby distribution that minimises the KL-divergence.

Since we've just observed $m$ agents in state $i$ for each $(i,m)\in\Omega$ we can define the $\Delta$ of the approximation to be
\begin{equation}
\Delta_i =
\begin{cases}
m & \text{if }(i,m)\in \Omega \\
0 & \text{otherwise}
\end{cases}
\label{deltaassimilate}
\end{equation}
given this, $\lambda_i$ is given by
\begin{equation}
\forall_i:  \lambda_i = \frac{ \emptyset \cdot e^{\sum_j a_j} a_i \left(\prod_{(j,m)\in\Omega} a_j^{\dag m}a_j^{m} Lg_{jr}\right)e^{tH}\psi_0}{\emptyset \cdot e^{\sum_j a_j} \left(\prod_{(j,m)\in\Omega} a_j^{\dag m}a_j^{m} Lg_{jr}\right)e^{tH}\psi_0}
  - \Delta_i
\label{lambdaassimilate}
\end{equation}

To calculate this, we need an algorithm to calculate expressions of the form
\[
y = \emptyset \cdot e^{\sum_j a_j} X e^{tH}\psi_0
\]
To increase numerical stability, we transform this to
\[
y = e^{-t}\emptyset \cdot e^{\sum_j a_j} X e^{t(I + H)}\psi_0
\]
where $I$ is the identity operator. Expanding the exponential in $H$ gives
\begin{equation}
y = e^{-t}\sum_{n=0}^\infty \frac{t^n}{n!} \emptyset \cdot e^{\sum_j a_j} Z_n \psi_0
\label{expansion}
\end{equation}
where $Z_n = X(I+H)^n$. However,
\[
Z_{n+1} = Z_n(I+H) = Z_n + HZ_n + [Z_n,H]
\]
but since $H$ is the Hamiltonian, and the Hamiltonian preserves probability mass,
\[
\emptyset \cdot e^{\sum_j a_j} HZ_n\psi_0=0
\]
for all $Z_n$ and $\psi_0$ so we can replace $Z$ by $Z'$ in equation \ref{expansion} where $Z'_0 = X$ and
\begin{equation}
Z'_{n+1} = Z'_n + [Z'_n,H]
\label{zrecurrence}
\end{equation}
without affecting the value of $y$.

This is very convenient computationally, because although $H$ is in general very large, $[Z'_n,H]$ is considerably smaller so we have a much more efficient way to calculate the terms of the expansion.

The computation can be made even more efficient by noting that
\[
\emptyset \cdot  e^{\sum_j a_j} a_i^\dag X = \emptyset \cdot  e^{\sum_j a_j} X
\]
for all $i$. So when calculating each $Z'_n$ we can use the commutation relations to remove all creation operators by moving them to the left, leaving only annihilation operators.

In practice, we only need to calculate the first few terms in the expansion in \ref{expansion} so we end up with
\begin{equation}
y = e^{-t}\sum_{n=0}^N \frac{t^n}{n!} \emptyset \cdot e^{\sum_j a_j} Z'_n \psi_0
\label{sumexpansion}
\end{equation}
The $Z'_n$ terms, can be calculated consecutively using equation \ref{zrecurrence} with
\[
Z'_0 = a_i \prod_{(j,m)\in\Omega} a_j^{\dag m}a_j^{m} Lg_{jr}
\]
for the numerator in equation \ref{lambdaassimilate} and
\[
Z'_0 = \prod_{(j,m)\in\Omega} a_j^{\dag m}a_j^{m} Lg_{jr}
\]
for the denominator.

We then just need to multiply $Z'_n$ by $\psi_0$ (i.e. the Deselby distribution from the previous assimilation cycle), convert to reduced form and sum the coefficients (in order to perform the $\emptyset \cdot e^{\sum_j a_j}$ operation)\footnote{We can also sum coefficients when working with a Deselby ground state since the sum of the coefficients of a Deselby distribution, expressed as an operator on $\emptyset$, is 1}.

Numerical experiments showed that, when calculating the terms in equation \ref{sumexpansion}, a good estimate of the error obtained by truncating the series at the $N^{th}$ term can be calculated as
\begin{equation}
\epsilon \approx e^{-t}\left( e^{\bar{x}} - \sum_{n=0}^N\frac{\bar{x}^n}{n!}\right)
\label{truncation}
\end{equation}
where
\[
\bar{x} = \frac{t}{N}\sum_{n=1}^N \left|\emptyset \cdot e^{\sum_j a_j} Z'_n \psi_0\right|^{\frac{1}{n}}
\]
This can be calculated along with the terms of the expansion until the error estimate falls below some desired threshold.

A final computational simplification can be made by noticing that when calculating the numerator of equation \ref{lambdaassimilate}, observations far away from the $i^{th}$ grid-square will have very little influence on its posterior (depending on $t$ and the speed of information flow through the ABM).

This intuition is formalised in Appendix \ref{separationderivation} where we show that if we let $\Phi(X)$ be the set of all states operated on by operator $X$, and let
\[
[^n X,H] = [...[[X,H],H]...,H]
\]
denote the $n$-fold application of a commutation, then if the observations in $\Omega$ can be partitioned into two sets $\Omega_A$ and $\Omega_B$ in such a way that
\[
\Phi\left(\sum_{n=0}^N[^na_i\prod_{(m,j)\in\Omega_A}a_j^{\dag m}a_j^m,H]\right) \cap \Phi\left(\sum_{n=0}^N[^n\prod_{(m,j)\in\Omega_B}a_j^{\dag m}a_j^m,H]\right) = \emptyset
\]
(where $\emptyset$ above is the empty set, not the ground state) we can factorise out the effect of all observations in $\Omega_B$ (to $N^{th}$-order accuracy) in both the numerator and denominator so that they cancel and we can effectively remove the observations in $\Omega_B$ (note that in the above equation we've removed the $Lg_{jr}$ operator as it has no effect under the $\Phi$ operation).

\section{Spatial predator-prey example model}

In order to provide a numerical demonstration of the above techniques, we simulated a 32x32 grid of agents. Each agent could be either a ``predator'' or a ``prey''. At any time a prey had a propensity to move to an adjacent grid-square (i.e. up, down, left or right), to die or to give birth to another prey on the an adjacent grid-square. Predators also had propensities to move to adjacent grid-squares or die, but in the presence of a prey on an adjacent grid-square, they also had a propensity to eat the prey and reproduce into an adjacent grid-square. Perhaps surprisingly, even a model with such a simple set of behaviours exhibits quite complex emergent properties that are difficult to predict. The rate per unit time of each behaviour is shown in table \ref{rates}.

\begin{table}
\begin{center}
\begin{tabular}{llc}
\hline
Aget type & Description & Rate per unit time\\
\hline
Prey & death &        0.1\\
 & reproduction &        0.15\\
 & movement &        1.0\\
 &&\\
Predator & death  &      0.1\\
 & predation/reproduction (per prey) &        0.5\\
 & movement &        1.0\\
\hline
\end{tabular}
\end{center}
\caption{The rates of each behaviour in the predator-prey model}
\label{rates}
\end{table}

A non-probabilistic, forward simulation of the ABM was performed in order to create simulated observation data. The initial state of the simulation was constructed by looping over all grid-squares and choosing the number of predators in that square from a Poisson distribution with rate parameter $\lambda = 0.03$ and the number of prey from a Poisson distribution with $\lambda = 0.06$. The model was then simulated forward in time and at time intervals of 0.5 units an observation of the state of the simulation was recorded. An observation was performed as follows: for each grid-square, with 0.02 probability, the number of prey was observed. If an observation was made, each prey in that grid-square was counted with a 0.9 probability. The same procedure was then repeated for predators.

Once the observations were generated, a sequence of data assimilation cycles were performed as described in equations \ref{deltaassimilate} and \ref{lambdaassimilate}, with a starting state of the Deselby ground state $D_\Lambda$ with $\lambda_i = 0.03$ for all $i$ representing a predator and $\lambda_i = 0.06$ for all $i$ representing a prey. The terms in the expansion were calculated until the expected truncation error, calculated using equation \ref{truncation} fell below 0.2\% of the sum of terms up that point.

\subsection{Results}

The mutual information between the assimilated state and the real state (where the real state is represented by a Dirac delta distribution) was calculated at the end of each assimilation cycle. In order to distinguish between the information accumulated by the data assimilation cycles and the information contained in the current time's observation and the prior at the start of the simulation, we also calculated a reference value equal to the mutual information between
\[
\frac{\left(\prod_{(j,m)\in\Omega_t} a^{\dag m}_ja_j^mLg_{ir}\right)D_\Lambda}{\emptyset \cdot e^{\sum_i a_i}\left(\prod_{(j,m)\in\Omega_t} a^{\dag m}_ja_j^mLg_{ir}\right)D_\Lambda}
\]
and the real state (i.e. the posterior given only the initial prior and the current window's observations).

\begin{figure}
\begin{center}
\includegraphics[width = 12cm]{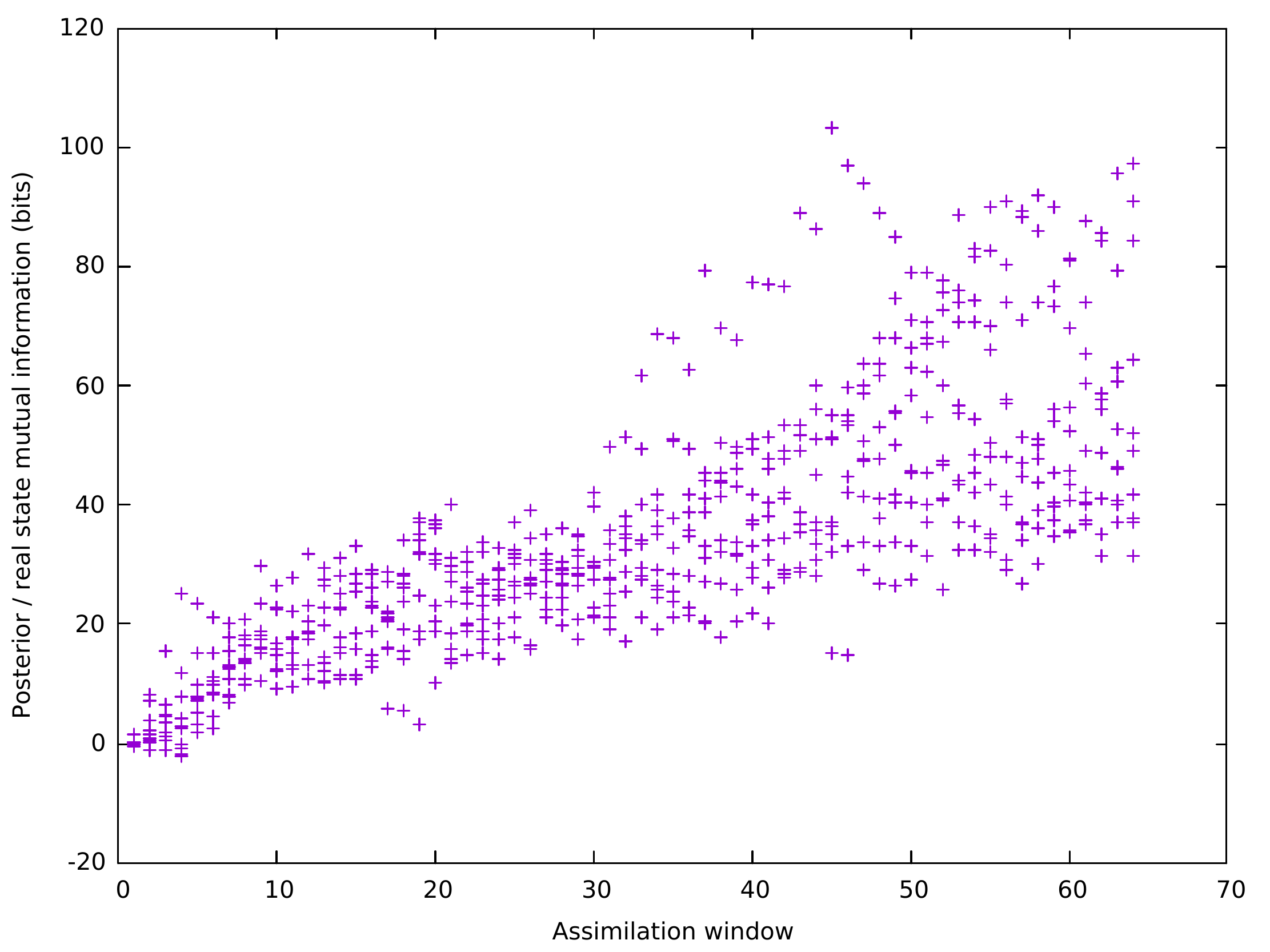}
\end{center}
\caption{Plot showing the increase of mutual information between the computed posterior distribution and the actual state for the first 64 assimilation windows of 10 simulations. The zero-information point is taken to be that of the posterior given the current window's observations and the prior at the start of the simulation.}
\label{information}
\end{figure}

Figure \ref{information} shows the difference between the mutual information of the assimilated state and the reference value for each assimilation cycle of 10 separate simulations. The upward trend in this plot shows that the assimilation cycle is successfully accumulating information about the real state.

\begin{figure}
\begin{center}
\includegraphics[width = 10cm]{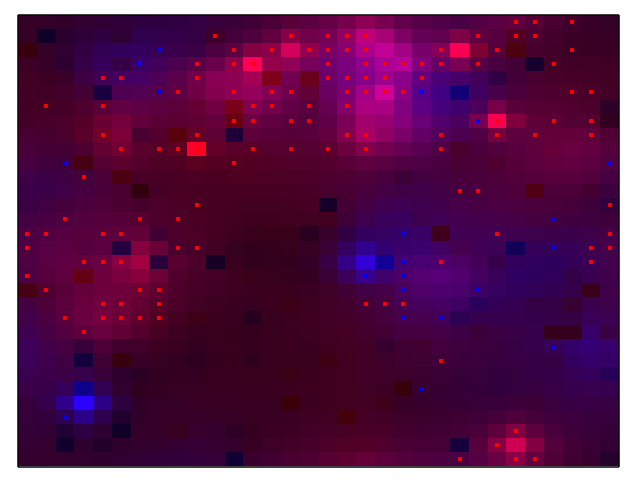}

\includegraphics[width = 10cm]{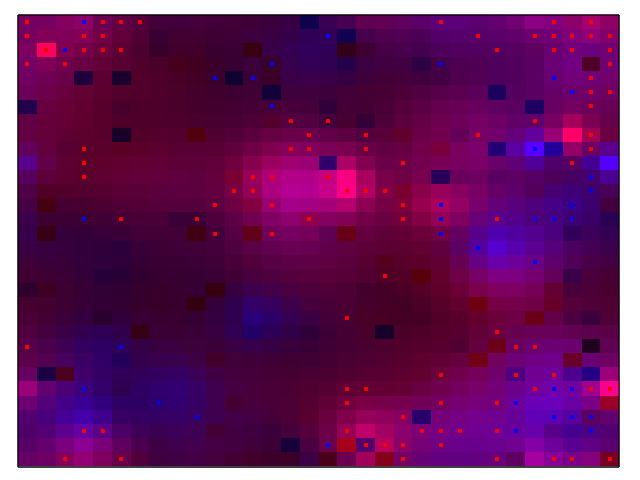}

\includegraphics[width = 10cm]{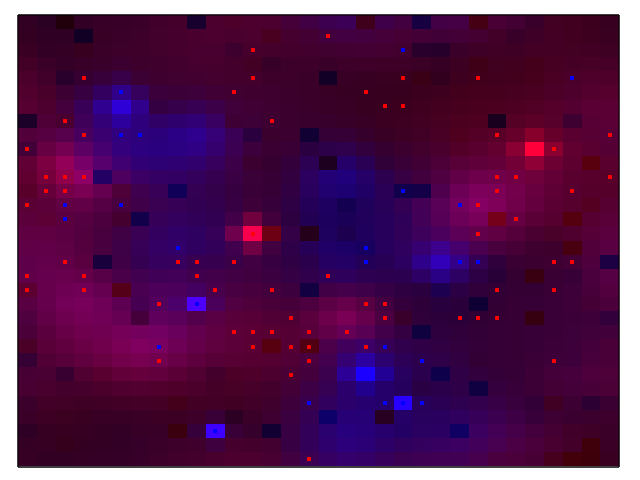}
\end{center}
\caption{Posterior, assimilated probability (background colour) plotted against actual agent positions (dots) for the highest (top), average (middle) and lowest (bottom) mutual information (out of 10 simulations) after 64 assimilation cycles. Red dots are real prey positions, blue dots are real predator positions. The background colour shows the posterior probability: higher intensity red corresponds to higher probability of prey, higher intensity of blue corresponds to higher probability of predator.}
\label{snapshots}
\end{figure}

Figure \ref{snapshots} shows snapshots of the real state of the non-probabilistic simulation superimposed on the assimilated state, for three of the simulations after 64 assimilation cycles. Inspection of these snapshots seems to suggest that the mutual information tends to be higher when agents are arranged in clusters rather than being more scattered, which is not surprising. Also, when a grid-square containing an agent is observed but (because of the 0.9 probability of observation) the number of observed agents happens to be zero, this will reduce the mutual information.

\section{Previous work}

Applying the methods of quantum field theory to other domains is not new; for example Dodd and Ferguson\cite{dodd2009many} suggest applying it to epidemiology, O'Dwyer and Green\cite{o2010field} apply it to simulations of biodiversity while Santos et.al.\cite{santos2015fock} applies it to enzymatic reations. Abarbanel\cite{abarbanel2013predicting} comes closest to our work in that he talks about quantum field theory in the context of data assimilation. However, to our knowledge, nobody has before developed quantum field theory to apply to such a general class of ABMs, nor have they developed it into an algorithm for data assimilation. Also, the various numerical strategies presented here that make the technique computationally tractable, including our presentation of the Deselby distribution is, we believe, new.

\section{Discussion and Conclusion}

This paper presents some early results of a promising new technique in data assimilation for agent based models. However, there's plenty of work left to be done and many opportunities for further development. For example, more research into flexible ways of approximating the state or computationally efficient ways of representing a state as a data structure, would be worthwhile. The example model presented here is very small and simple, the next step would be to apply this technique to larger and more complex models. In addition, in this paper we only consider agents with a discrete set of states, the extension to states including continuous variables is fairly straightforward and work on this is currently under way.

At present there are very few techniques that can be used to successfully perform data assimilation in agent based models. We believe the results presented here provide a significant first step towards filling that gap and hope that it will form the foundation of a fruitful approach to the problem.

The code used to create the results in this paper can be found at \href{https://github.com/deselby-research/ProbabilisticABM}{https://github.com/deselby-research/ProbabilisticABM}, where you'll also find a library of code we're developing in order to experiment with new techniques in this area.

\bibliographystyle{unsrt}  
\bibliography{references}

\newpage
\appendix
\section{Appendix: Binomial observation}
\label{binomialderivation}

The product of a Deselby distribution and a binomial distribution is given by
\[
P(k|m) \propto {k \choose m}r^m(1-r)^{k-m}\frac{(k)_\Delta \lambda^{k-\Delta} e^{-\lambda}}{k!}
\]
rearranging and dropping constants
\begin{equation}
P(k|m) \propto (1-r)^\Delta \, (k)_m \, \frac{(k)_\Delta ((1-r)\lambda)^{k-\Delta} e^{-(1-r)\lambda}}{k!}
\label{posterior}
\end{equation}
expanding the product of falling factorials
\[
P(k|m) = (1-r)^\Delta \sum_{q=0}^{\min(m,\Delta)} \frac{m!\Delta!\left((1-r)\lambda\right)^{(m-q)}}{q!(m-q)!(\Delta-q)!} \frac{(k)_{m + \Delta - q} ((1-r)\lambda)^{k-(m + \Delta - q)} e^{-(1-r)\lambda}}{k!}
\]
Expressing this as an operator on $\emptyset$
\[
P \propto \sum_{k=0}^\infty (1-r)^\Delta \sum_{q=0}^{\min(m,\Delta)} \frac{m!\Delta!\left((1-r)\lambda\right)^{(m-q)}}{q!(m-q)!(\Delta-q)!} \frac{(k)_{m + \Delta - q} ((1-r)\lambda)^{k-(m + \Delta - q)} e^{-(1-r)\lambda}}{k!}a^{\dag k}\emptyset
\]
so
\[
P \propto (1-r)^\Delta \sum_{q=0}^{\min(m,\Delta)} \frac{m!\Delta!\left((1-r)\lambda\right)^{(m-q)}}{q!(m-q)!(\Delta-q)!} a^{\dag (m+\Delta-q)}D_{i((1-r)\lambda)\,0}
\]
But, from the commutation relations, we have the identity
\[
a^m a^{\dag \Delta} = \sum_{q=0}^{\min(m,\Delta)} \frac{m!\Delta!}{q!(m-q)!(\Delta-q)!} a^{\dag \Delta-q}a^{m-q}
\]
so
\[
P \propto (1-r)^\Delta a^{\dag m}a^m a^{\dag\Delta}D_{((1-r)\lambda)\,0}
\]
Let $Lg_r$ be an operator that has the properties
\begin{equation}
[Lg_r, a^\dag] = -ra^\dag Lg_r
\end{equation}
\begin{equation}
[Lg_r, a] = \frac{ra}{1-r} Lg_r
\end{equation}
and
\[
Lg_rD_{\lambda0} = D_{(1-r)\lambda0}
\]
then
\[
Lg_ra^{\dag \Delta} = (1-r)^\Delta a^{\dag \Delta}Lg_r
\]
So
\[
P \propto  a_i^{\dag m}a_i^m Lg_{ir} D_{i\lambda\Delta}
\]

\newpage
\section{Appendix: Factorisation of compound observations}
\label{separationderivation}

Consider an expression of the form
\[
\emptyset\cdot e^{\sum_j a_j} ABe^{tH} = \sum_{n=0}^\infty\frac{t^n}{n!}\emptyset\cdot e^{\sum_j a_j}ABH^n
\]
where $A$, $B$ and $H$ are operators, and $H$ is a Hamiltonian.

Due to the form of a Hamiltonian, $\emptyset\cdot e^{\sum_j a_j}HX = 0$ for all X, so
\[
\emptyset\cdot e^{\sum_j a_j}ABH^n = \emptyset\cdot e^{\sum_j a_j}[^n AB, H]
\]
Where we introduce the notation
\[
[^n X,H] = [...[[X,H],H]...,H]
\]
to denote the $n$-fold application of a commutation.

So
\[
\emptyset\cdot e^{\sum_j a_j} ABe^{tH} = \sum_{n=0}^\infty\frac{t^n}{n!}\emptyset\cdot e^{\sum_j a_j}[^nAB,H]
\]
Now let
\[
C_n(A,B) = \sum_{m=0}^n {n \choose m}[^mA,H][^{n-m}B,H]
\]
so
\[
[C_n(A,B),H] = \sum_{m=0}^n {n \choose m}\left([^mA,H][^{n-m+1}B,H] + [^{m+1}A,H][^{n-m}B,H]\right)
\]
\[
 = \sum_{m=0}^n {n \choose m}[^mA,H][^{n+1-m}B,H] + \sum_{m'=1}^{n+1}{n \choose m'-1}[^{m'}A,H][^{n+1-m'}B,H]
\]
\[
 = \sum_{m=0}^{n+1} {n+1 \choose m}\frac{n+1-m}{n+1}[^mA,H][^{n+1-m}B,H] + \sum_{m'=0}^{n+1}{n+1 \choose m'}\frac{m'}{n+1}[^{m'}A,H][^{n+1-m'}B,H]
\]
\[
= \sum_{m=0}^{n+1} {n+1 \choose m}[^mA,H][^{n+1-m}B,H]
\]
So
\[
[C_n(A,B),H] = C_{n+1}(A,B)
\]
But
\[
[AB,H] = A[B,H] + [A,H]B = \sum_{m=0}^1 {1 \choose m}[^mA,H][^{1-m}B,H] = C_1(A,B)
\]
so
\[
[^n AB,H] = \sum_{m=0}^n {n \choose m}[^mA,H][^{n-m}B,H]
\]
So
\[
\sum_{n=0}^\infty \frac{[^nAB,H]t^n}{n!}
= \sum_{n=0}^\infty \sum_{m=0}^n {n \choose m} \frac{[^mA,H][^{n-m}B,H]t^n}{n!}
\]
\[
= \sum_{n=0}^\infty \sum_{m=0}^n \frac{[^mA,H]t^m[^{n-m}B,H]t^{n-m}}{m!(n-m!)}
= \sum_{j=0}^\infty \frac{[^j A,H]t^j}{j!} \sum_{k=0}^\infty \frac{[^k B,H]t^k}{k!}
\]
So
\[
\emptyset\cdot e^{\sum_j a_j} ABe^{tH}\psi D_\Lambda \approx \emptyset\cdot e^{\sum_j a_j}\sum_{j=0}^N \frac{[^j A,H]t^j}{j!} \sum_{k=0}^N \frac{[^k B,H]t^k}{k!}\psi D_\Lambda
\]
where $\psi$ is a sequence of creation operators and $D_\Lambda$ is the Deselby ground state.

If we let $\Phi(X)$ be the set of all states operated on by operator $X$, then if
\begin{equation}
\Phi\left(\sum_{n=0}^N[^nA,H]\right) \cap \Phi\left(\sum_{n=0}^N[^nB,H]\right) = \emptyset
\label{nointersect}
\end{equation}
where $\emptyset$ above is the empty set (not the ground state) then $\psi$ can easily be factorised into $\psi_A\psi_B$ such that
\[
\emptyset\cdot e^{\sum_j a_j}\sum_{j=0}^N \frac{[^j A,H]t^j}{j!} \sum_{k=0}^N \frac{[^k B,H]t^k}{k!}\psi D_\Lambda = \emptyset\cdot e^{\sum_j a_j}\sum_{j=0}^N \frac{[^j A,H]t^j}{j!}\psi_A \sum_{k=0}^N \frac{[^k B,H]t^k}{k!}\psi_B D_\Lambda
\]

But if $X$ consists only of annihilation operators (which we can arrange by using commutation relations to move all creation operators to the left and then removing them), then
\[
\emptyset\cdot e^{\sum_j a_j}XYD_0 = \left(\emptyset\cdot e^{\sum_j a_j}XD_0\right) \left(\emptyset\cdot e^{\sum_j a_j}YD_0\right) + \emptyset\cdot e^{\sum_j a_j}[X,Y]D_0
\]
but if equation \ref{nointersect} is satisfied then
\[
\left[\sum_{n=0}^N[^nA,H]\psi_A, \sum_{n=0}^N[^nB,H]\psi_B\right] = 0
\]
irrespective of whether we strip the first term of creation operators, so, if \ref{nointersect} is satisfied
\[
\emptyset\cdot e^{\sum_j a_j} ABe^{tH}\psi D_\Lambda \approx \left(\emptyset\cdot e^{\sum_j a_j}\sum_{j=0}^N \frac{[^j A,H]t^j}{j!}\psi D_\Lambda\right) \left(\emptyset\cdot e^{\sum_j a_j}\sum_{k=0}^N \frac{[^k B,H]t^k}{k!}\psi D_\Lambda\right)
\]
where we've removed the factorisation of $\psi$ as it doesn't affect the sum.

\end{document}